\documentclass[a4paper,11pt]{article}
\usepackage{pos}

\title{The Use of the Signal at an Optimal Distance from the Shower Core as a Surrogate for Shower Size}
\ShortTitle{Optimal distance and shower size}

\author[a]{Q. Luce}
\author[a]{D. Schmidt}
\author[b]{O. Deligny}
\author[b]{I. Lhenry-Yvon}
\author[c]{M. Roth}
\author[d]{A.A. Watson}

\affiliation[a]{Institute for Experimental Particle Physics, Karlsruhe Institute of Technology (KIT), Karlsruhe, Germany}
\affiliation[b]{CNRS/IN2P3, IJCLab, Université Paris-Saclay, Orsay, France}
\affiliation[c]{Institute for Astroparticle Physics, Karlsruhe Institute of Technology (KIT), Karlsruhe, Germany}
\affiliation[d]{School of Physics and Astronomy, University of Leeds, LS2 9JT, UK}

\emailAdd{quentin.luce@kit.edu}
\emailAdd{david.schmidt@kit.edu}

\abstract{When analysing data from air-shower arrays, it has become common practice to use the signal at a considerable distance from the shower axis ($r_\text{opt}$) as a surrogate for the size of the shower.
This signal, $S(r_\text{opt}$), can then be related to the primary energy in a variety of ways.
After a brief review of the reasons behind the introduction of $r_\text{opt}$ laid out in a seminal paper by Hillas in 1969, it will be shown that $r_\text{opt}$, is a more effective tool when detectors are laid out on a triangular grid than when detectors are deployed on a square grid.
This result may have implications for explaining the differences between the flux observed by the Auger and Telescope collaborations above 10\,EeV and should be kept in mind when designing new shower arrays.}

\FullConference{%
   27th European Cosmic Ray Symposium - ECRS \\
   25-29 July 2022 \\
   Nijmegen, the Netherlands 
}

\begin{document}
\maketitle

\section{Introduction: Why an optimal distance?}

Ground based experiments dedicated to the observation of ultra-high energy cosmic-rays (UHECRs) at energies above 10$^{18}$\,eV are covering larger and larger areas of detection.
In the 1970s, the Volcano Ranch and Haverah Park~\cite{HaverahParkSpectrum} experiments respectively covered areas of $\sim$8.1\,km$^2$ and 12\,km$^2$. The Pierre Auger Observatory~\cite{AugerSpectrum2020} and Telescope Array~\cite{TASpectrum2019}, which are currently collecting data, boast respective areas of $\sim$3,000\,km$^2$ and $\sim$800\,km$^2$.
The next generation observatory is planning for an area a factor 10 larger than the present experiments. 
Building arrays covering such large areas at a reasonable cost implies spacing detectors (usually water-Cherenkov detectors or scintillators) at distances several times larger than the Molière radius, which marks the area containing $\sim$90\% of the ionizing particles of an extensive air-shower.
Consequently, the distribution of these particles as a function of the distance from the shower axis, i.e. the lateral profile of the shower, is sampled by only a handful of detectors.
Moreover, the densities of particles have to be extrapolated close to the core, where the saturation of the detectors results in inaccurate measurements.
In such a configuration, the parametrisation of the lateral distribution function (LDF), which models the fall-off of the density of particles with distance from the shower axis, and consequently the determination of the total number of particles in the air-shower, becomes a difficult task.

For example, in the early days of the analysis of events observed at Haverah Park, when only four stations had been deployed, the signal recorded was assumed to follow $r^{-n}$, where $r$ is the distance from the shower axis. 
The slope parameter $n$ suffers from shower-to-shower fluctuations and varies with energy. 
The energy of the primary cosmic-ray was determined by integrating the signal which would have been observed from 100$\,$m to 1000$\,$m from the axis of the shower. 
Any variation of $n$ therefore translated into a variation in the reconstructed energy, which resulted in large systematic uncertainties.
A modification of $n$ by 0.6 modified the reconstructed energy by 70\%~\cite{Hillas1970}.
To overcome this issue, Hillas analysed 50 showers observed by the Haverah Park experiment and found that the density of particles at 500$\,$m from the axis was altered by only $\sim$12\% for the same variation of $n$. 
The signal at this \emph{optimal} distance, appropriate for the array of water-Cherenkov detectors at Haverah Park, where three detectors were placed equidistant on the circumference of a circle of radius of 500$\,$m with a fourth at the center, was then used as a more accurate estimator of the shower size and by extension, the energy of the primary. 
Later, with the final geometry of the array, the density of particles at a distance of 600$\,$m was chosen as a surrogate for the shower size.
This method opened a new era for the reconstruction of showers observed by ground-based arrays such as AGASA in the 1980s or the Pierre Auger Observatory (Auger) and Telescope Array (TA) today. 

The form of LDF used by these experiments is derived from \cite{Kamata1958} and \cite{Greisen1960}, and is a combination of power-laws with logarithmic slopes governing the fall-off of the signal in different ranges of distances from the shower axis.
Considering the characteristics intrinsic to a specific experiment (location and geometry of the array, choice of detectors, etc.), each experiment must determine a parametrisation of the LDF and an optimal distance appropriate for the characteristics of its array to reconstruct the data as accurately as possible.
Such studies have been performed by AGASA in \cite{Dai1988} for a square grid of scintillator detectors with a spacing of 1000$\,$m.
These results are currently also used to reconstruct events observed by TA.
In the analysis of showers observed by AGASA (where the layout was complex for logistical reasons), a distance around $\sim$600$\,$m was found to be the optimal choice to estimate the shower size.
However, the optimal distance was found to be dependent on the energy of the primary cosmic ray and the inclination of the shower\footnote{To second order, a dependency with the mass was also observed.}.
For a triangular grid of water-Cherenkov detectors spaced by 1500$\,$m, as used by Auger, the distance of 1000$\,$m was found to be the optimal choice to reconstruct the shower size considering events for which no detectors saturate.
When saturation occurs, the optimal distance shifts towards the closest distances at which detectors are not saturated. 
It was shown for such an array that the optimal distance is dependent on the distance between the detectors, but contrary to the results from AGASA, no dependencies on primary energy or shower inclination were observed~\cite{Newton2007}. 

In this contribution, we derive the optimal distance, using simulations that mimic the Auger and TA projects and explore if a choice of a \emph{non-optimal} distance, i.e. distance at which the fluctuations of the LDF are not minimal, can introduce a bias in the determination of primary energy and, as a consequence, in the reconstructed spectrum of UHECRs.

\section{Determination of the optimal distance}

The optimal distance, $r_\text{opt}$, as defined by Hillas et al.~\cite{Hillas1970, Hillas1971} corresponds to the distance at which the uncertainties of the lateral profile due to the lack of knowledge of the true shape of the LDF, minimise the uncertainties in the signal size. 
So, for a particular LDF, computing $r_\text{opt}$ consists in reconstructing an event a large number of times, each time with a different value for the logarithmic slope, and then scanning the relative deviations of the signal, $\sigma_{S}(r) / S(r)$, as a function of the distance. 
The distance at which $\sigma_{S}(r) / S(r)$ is minimal corresponds to $r_\text{opt}$.

The study has been performed using simulated events mimicking the response to a shower of the Pierre Auger Observatory or Telescope Array.
Thus simulated showers have been thrown on a triangular grid of water-Cherenkov detectors, separated by 1500$\,$m, called \emph{Auger-1500}.
In parallel, a second set of events has been simulated considering scintillators on a square grid spaced at 1200$\,$m, \emph{TA-1200}. 
Both sets presented here have been performed using simulation softwares developed by the Pierre Auger Collaboration~\cite{Argiro2007}. 
Cross-checks with a toy model Monte-Carlo for which the signal of a detector is sampled from a fixed LDF have been conducted, and lead to the same conclusions than the ones presented in this paper.

The tests were performed using 120 proton-induced air-shower simulations at energies of $10^{18.5}$, $10^{19.0}$, $10^{19.5}$ and $10^{20}\,$eV and, at inclinations of 0$^\circ$, 32$^\circ$ and 48$^\circ$ adopting EPOS-LHC and QGSJet-II.04 as hadronic model.
Each event, whether it contains a saturated station or not, is then reconstructed 100 times following the likelihood-procedures described in \cite{SDRec2020} or in \cite{IvanovThesis2012} for Auger-1500 or TA-1200 respectively.
In the case of Auger-1500 simulations, the LDF used to reconstruct the lateral profile is

\begin{equation}
    S_\text{Auger}(r) = S(r_\text{opt}) \left(\dfrac{r}{r_\text{opt}}\right)^{-\beta(\theta, S(r_\text{opt}))} \left(\dfrac{r+700}{r_\text{opt}+700}\right)^{-\beta(\theta, S(r_\text{opt}))},
    \label{eq:AugerLDF}
\end{equation}
where $r$ is the distance to the shower axis, $r_\text{opt} = 1000\,$m and the slope $\beta$ is modelled as a polynomial function of the logarithm of the estimator of the shower size $S(r_\text{opt})$ and of $\cos\theta$. 
To give an idea, the values of $\beta$ are decreasing with the shower size and with the inclination of the shower from $\sim2.6$ to $\sim1.5$.
Thus for each of the 100 reconstructions, the slope of the LDF is drawn according to a Gaussian distribution centered on the parametrisation of $\beta$ and with a standard deviation $\sigma_\beta = 0.428\exp(-0.406\lg(S(r_\text{opt})))$.
With the increase of the multiplicity of the events with energy, the fluctuations of $\beta$ fall-off rapidly. 
The relative fluctuations of the slope are then evolving with the inclination of the shower from 12 to 19\% at $10^{18.5}\,$eV and from 6 to 8\% at at $10^{20}\,$eV.

For the TA-1200 set, mimicking the TA scintillators array, the LDF used is the one parametrised by the AGASA Collaboration~\cite{Yoshida1994} and written as

\begin{equation}
    S_\text{TA}(r) = S(r_\text{opt})\left(\dfrac{r}{r_\text{opt}}\right)^{-\alpha} \left(\dfrac{r+91.6}{r_\text{opt}+91.6}\right)^{-(\eta(\theta)-\alpha)}\left(\dfrac{r^2+1000^2}{r_\text{opt}^2+1000^2} \right)^{-\gamma},
    \label{eq:TALDF}
\end{equation}
where $\alpha = 1.2$, $\gamma = 0.6$, $r_\text{opt} = 800\,$m and with $\eta = 3.97 - 1.79(\sec\theta-1)$. Each TA-1200 events is thus reconstructed 100 times with a slope $\eta$ drawn according to a Gaussian distribution centered on $\eta(\theta)$ and with a standard deviation, $\sigma_\eta$, fixed to 0.187 following the work reported in \cite{Linsley1977}.
It is thus driven relative fluctuations of $\eta$ with the inclination of the shower from 5 to 8\% (values comparable to the relative fluctuations at the highest energies of the slope $\beta$ in Eq.~\ref{eq:AugerLDF}).
In each reconstruction, three parameters are fitted, the coordinates of the shower core on ground and the estimator of the shower size $S(r_\text{opt})$.
The logarithm slope $\beta$ in Eq.~\ref{eq:AugerLDF} and $\eta$ in Eq.~\ref{eq:TALDF} are not set as free parameters because in the observed data of Auger or TA, the bulk of events contain of too few detectors to reconstruct all the characteristics of the LDF at the same time.

The optimal distance is constrained by the presence of detectors around it. 
In addition to the effect of the geometry of the array that fixes the position of the detectors, the saturation of one or more of the closest detectors would lead to shift the optimal distance towards the closest non-saturated detector as shown in \cite{Newton2007}. 
To simplify the discussion, the events containing a saturated detector are discarded from the following presentation of results.

\begin{figure}[t]
\centering\includegraphics[width=0.49\textwidth]{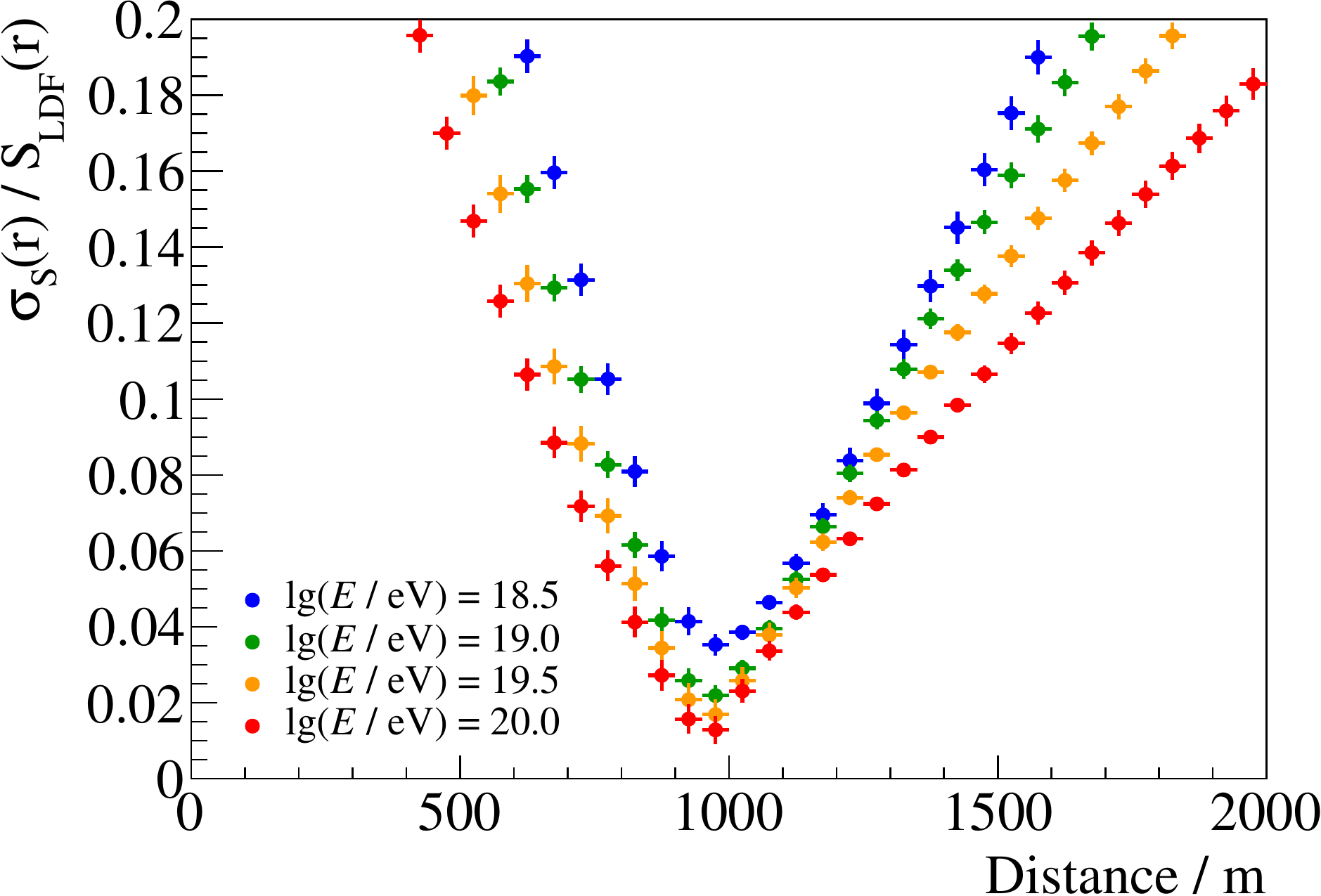}
\centering\includegraphics[width=0.49\textwidth]{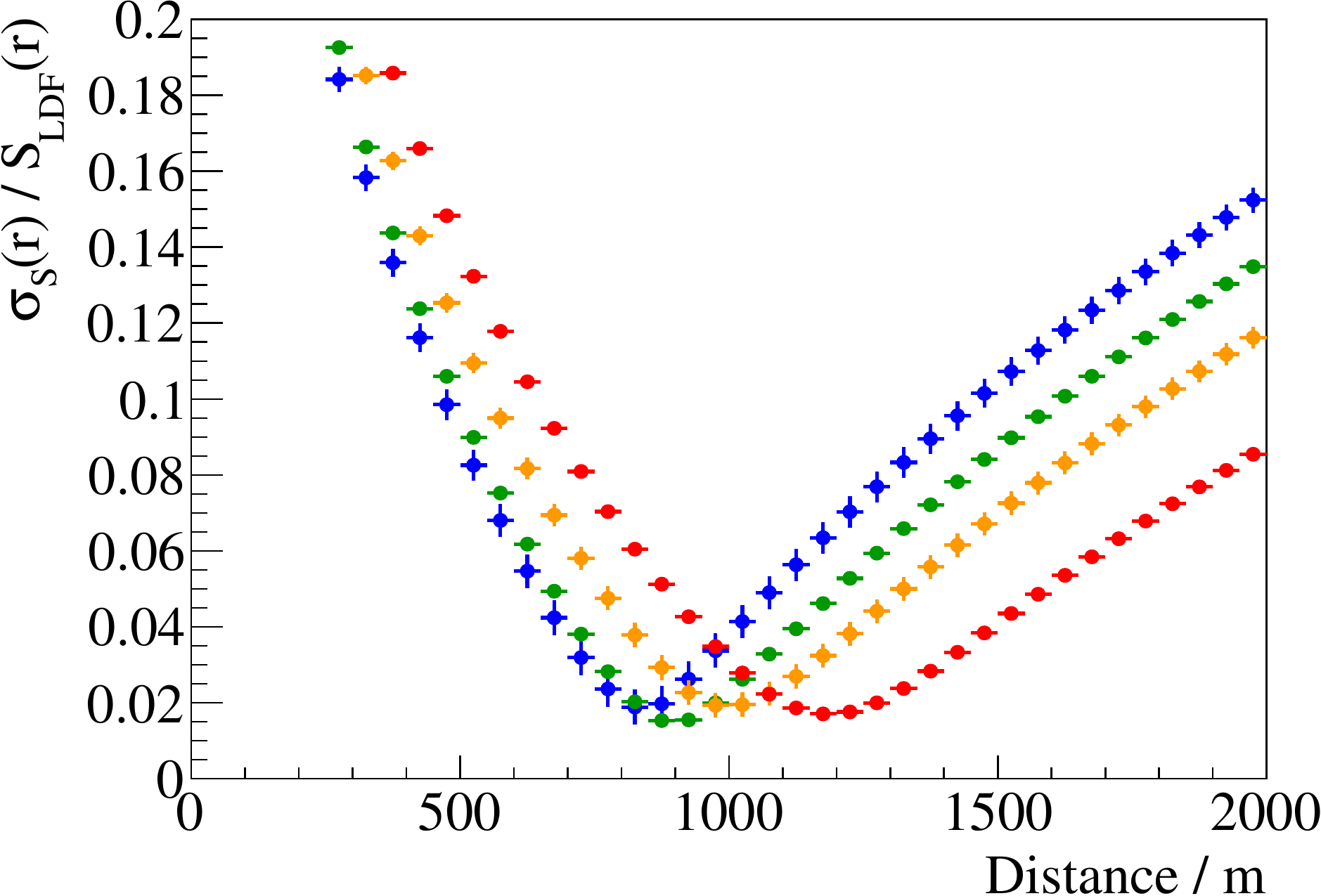}
\centering\includegraphics[width=0.49\textwidth]{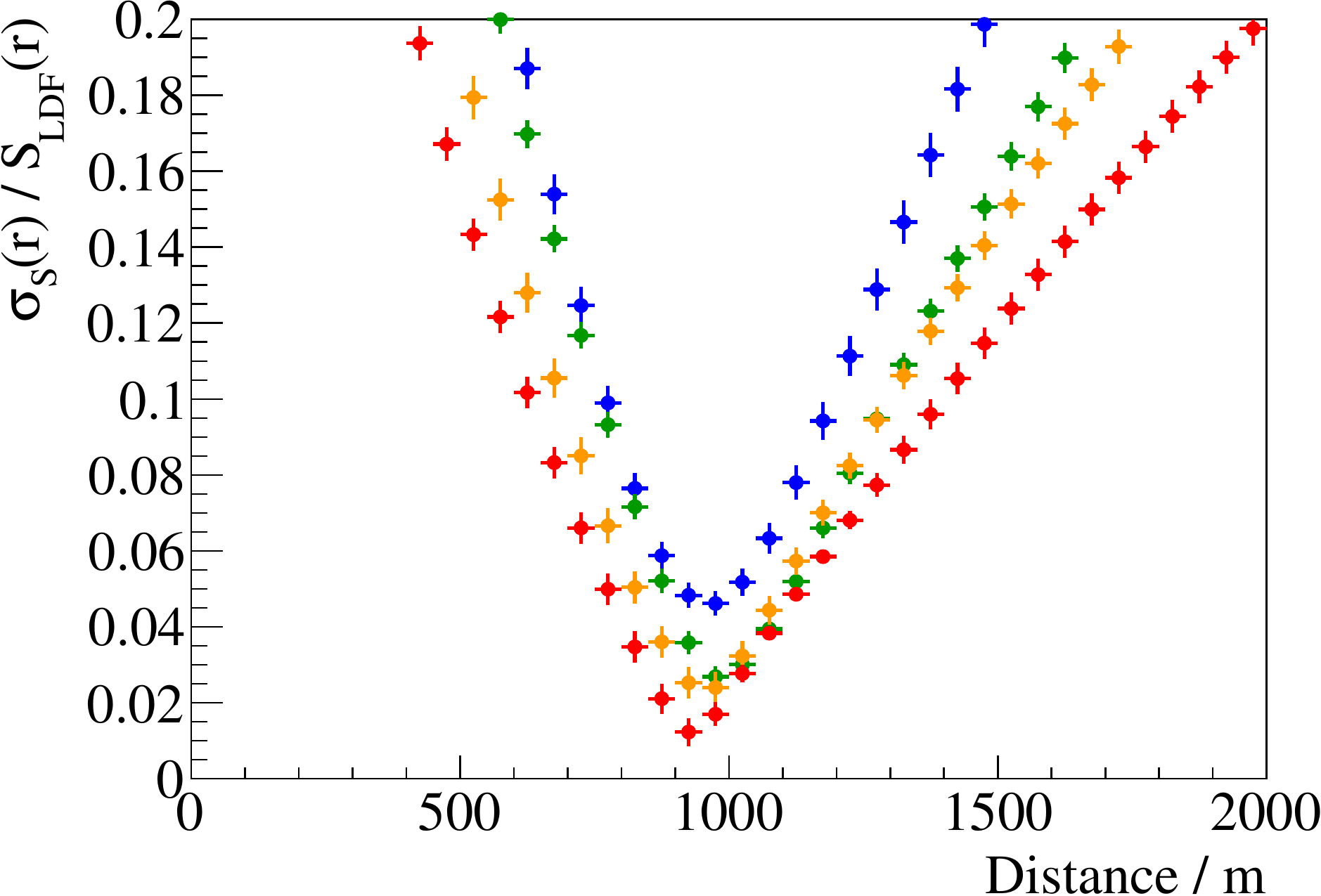}
\centering\includegraphics[width=0.49\textwidth]{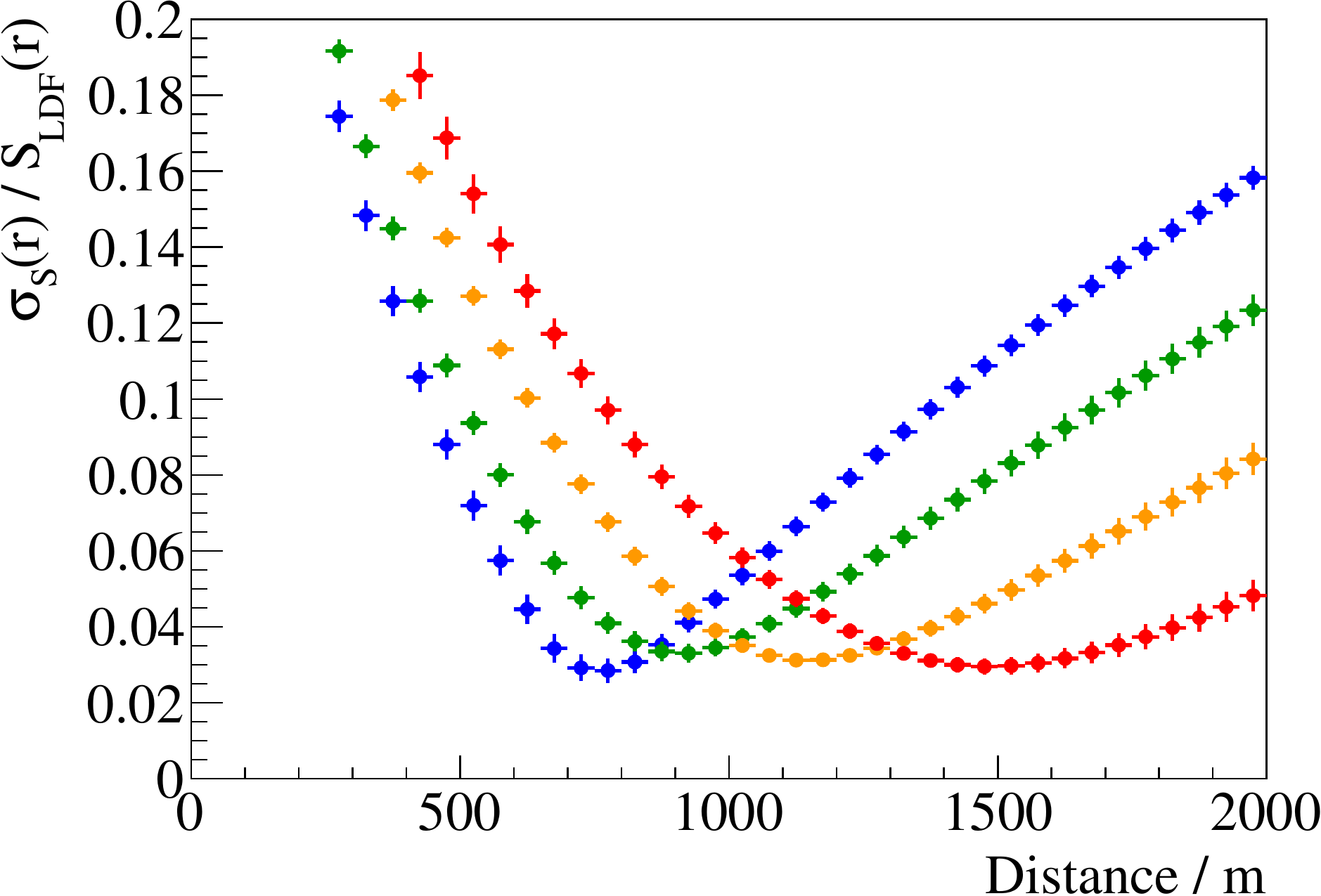}
\caption{\small{Relative fluctuations of the signal $\sigma_S(r)/S(r)$ as a function of distance for events simulated with QGSJetII.04 on the Auger-1500 (left) and on the TA-1200 (right) geometries. Plots in the top and bottom rows correspond to zenith angles of 0$^\circ$ and 48$^\circ$.}}
\label{fig:ropt}
\end{figure}

The relative fluctuations of the signal $\sigma_{S}(r) / S(r)$ are shown in Fig.~\ref{fig:ropt}. 
The values of $r_\text{opt}$ of the Auger-1500 configuration confirm the results reported in \cite{Newton2007} with an optimal distance around 1000$\,$m and independent of the energy of the primary or the inclination of the shower.
The results for the TA-1200 configuration is, however, quite different compared to the Auger-1500 configuration. 
The values of $r_\text{opt}$ are found to be dependent on the energy of the primary and on the inclination of the shower as stated in \cite{Dai1988}.
For all inclinations of the shower, at 800$\,$m which is the value of $r_\text{opt}$ used in Eq.~\ref{eq:TALDF}, the fluctuations of the signal increase from $\simeq2\%$ at $10^{18.5}$ eV, to $\simeq10\%$ at $10^{20.0}$ eV.
But at the highest energies considered, the optimal distance increases from $\simeq 1200\,$m for vertical events to $\simeq 1500\,$m at 48$^\circ$.
For both configurations, no dependencies on the mass or the hadronic model used have been observed.
Interestingly, in events for which scintillators replace the water-Cherenkov detectors on the triangular array and, reconstructed with the TA procedure, the energy and zenith dependencies of $r_\text{opt}$ are also observed while the values of $r_\text{opt}$ are modified by only few percents.

\section{Bias, resolution and spectrum}

While the choice of an optimal distance for Auger-1500 fixed at $\sim1000\,$m is confirmed, the choice of 800$\,$m for TA-1200 raises questions. 
The calibration in energy is performed assuming that, for any choice of LDF, the estimator of the shower size does not present strong non-linearities that cannot be absorbed by calibration procedure. 
If this assumption is not verified, then the energy will inherit the non-linearities of its estimator. 

\begin{figure}[t]
\centering\includegraphics[width=0.49\textwidth]{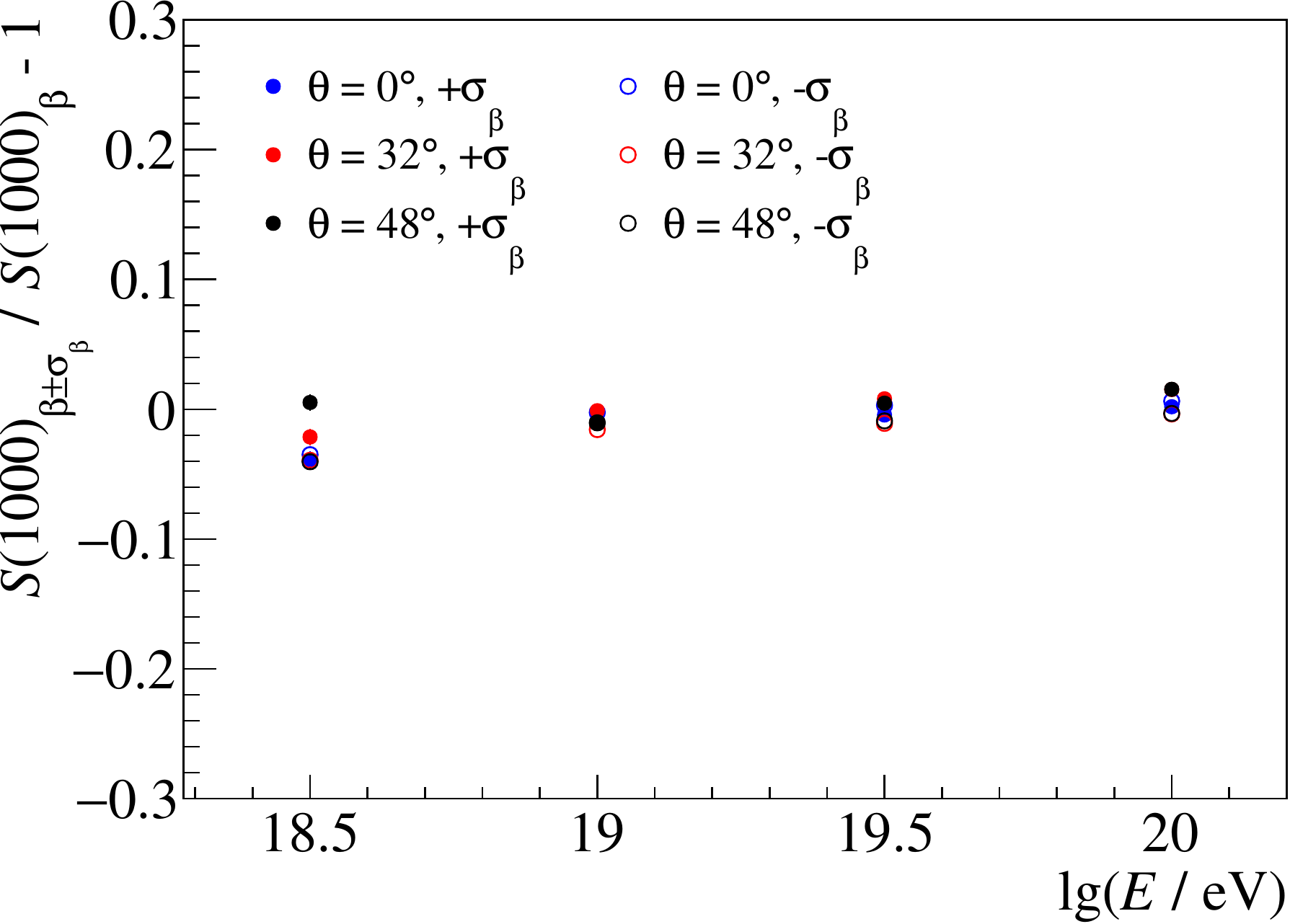}
\centering\includegraphics[width=0.49\textwidth]{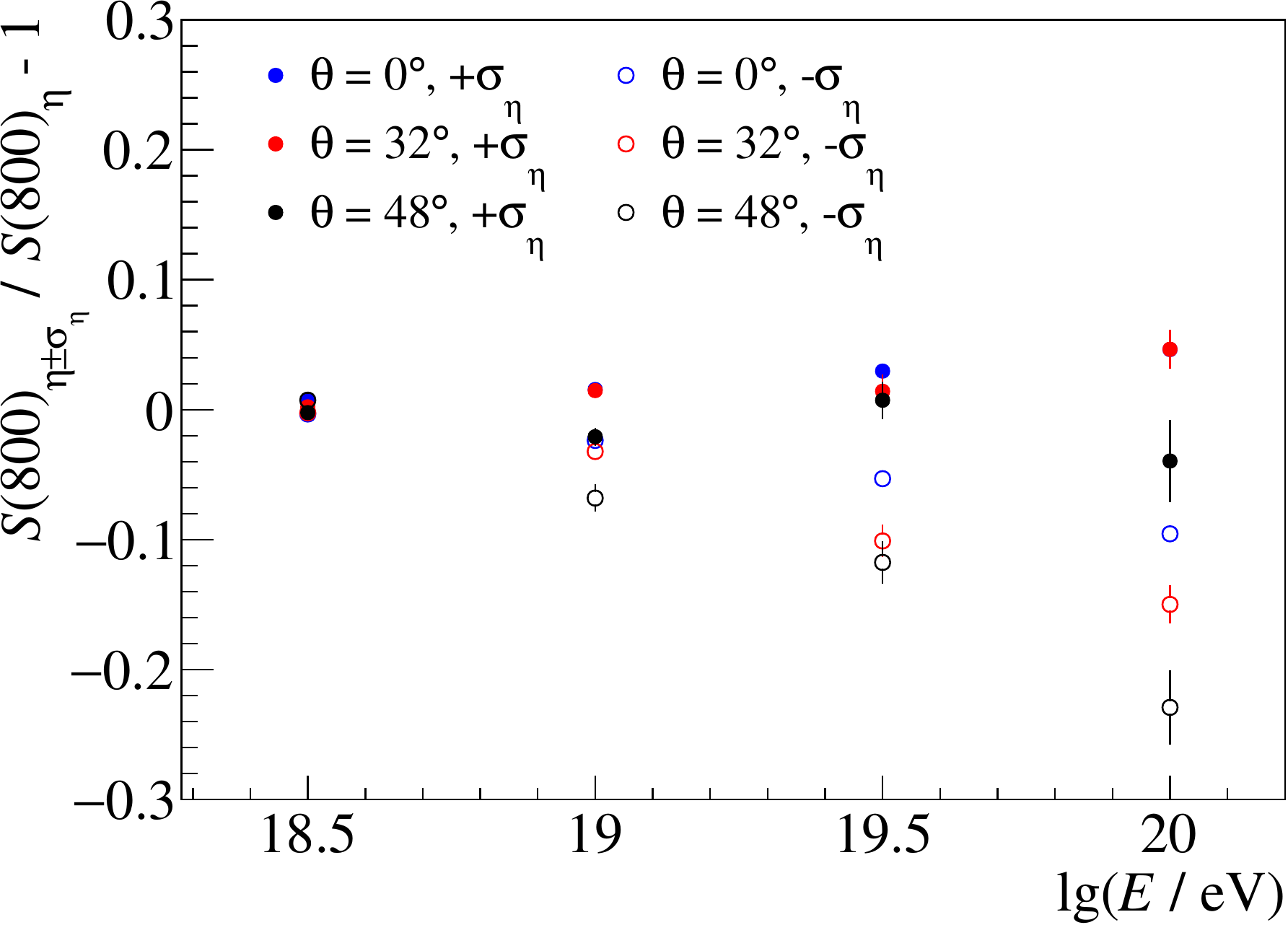}
\caption{\small Residuals in $S(r_\text{opt})$ for two specific departures of the logarithmic slope of the LDF with respect to the reference one. The results are shown for Auger-1500 (\emph{left}) and TA-1200 (\emph{right}).}
\label{fig:dsref}
\end{figure}

To test whether a specific choice of the LDF, i.e. a specific choice of the slope of the LDF could lead to the introduction of a bias in the energy, the simulated events for both Auger-1500 and TA-1200 sets have been reconstructed increasing the value of the slope of the LDF by $\pm \sigma$.
The residuals of $S(r_\text{ropt})_{\pm\sigma} / S(r_\text{ropt})$ are shown in Fig.~\ref{fig:dsref}. 
In the case of the Auger-1500 configuration, the fluctuations of the estimator of the shower size from the mean LDF are below $\simeq5\%$ and are the same for an increase or decrease of the slope $\beta$ by $\pm 1 \sigma_\beta$.
However, for the TA-1200 configuration, the non-linearities reaches $\simeq10\%$ per decade and is dependent on the specific LDF chosen, i.e. whether the slope $\eta$ is increased or decreased by $\sigma_\eta$.
In that case, important non-linearities of the energy estimator are expected and their impact on the energy spectrum has to be evaluated. 

\begin{figure}[t]
\centering\includegraphics[width=0.49\textwidth]{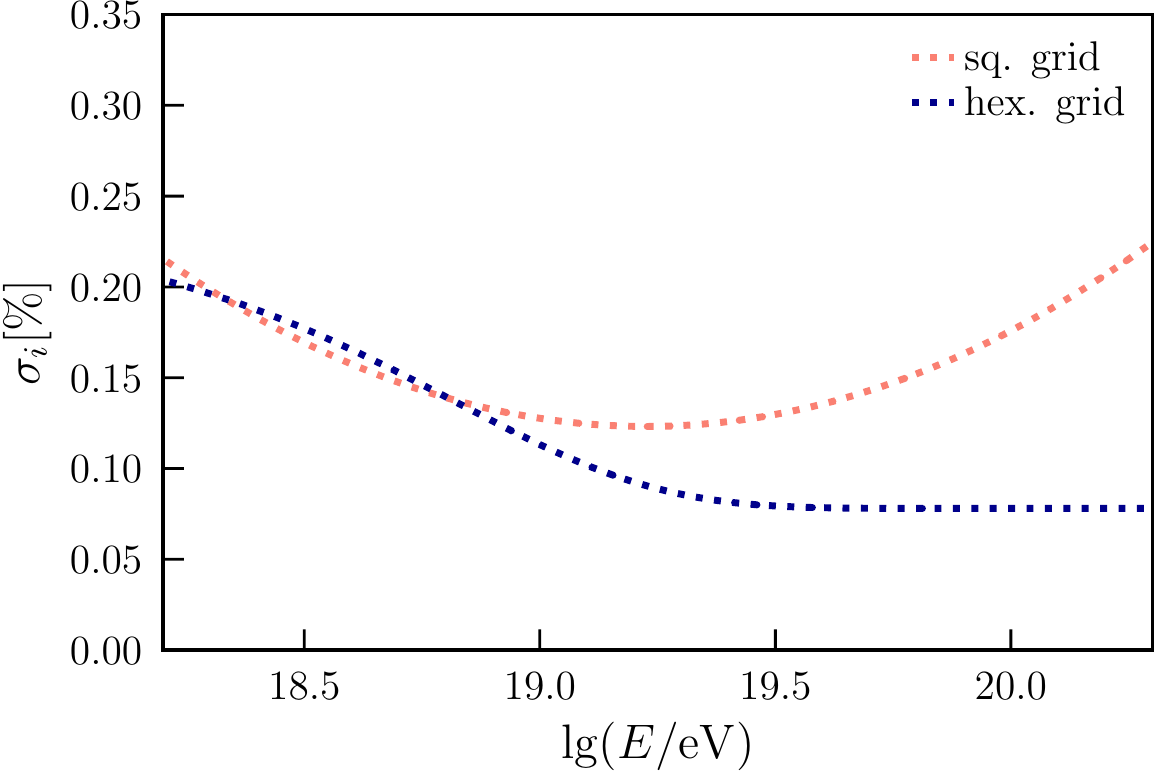}
\centering\includegraphics[width=0.49\textwidth]{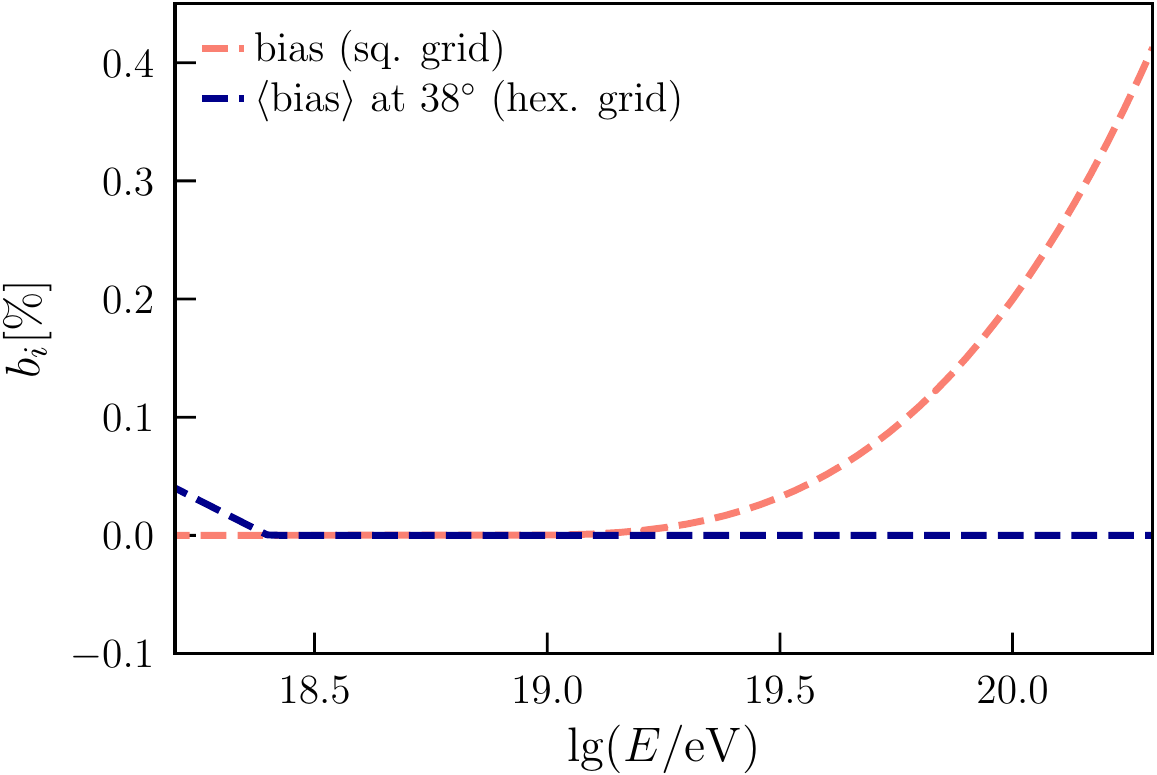}
\caption{\small \emph{Left}: Relative resolution of the Auger-1500 (blue dotted line) and TA-1200 (red dotted line). \emph{Right}: Relative bias of the Auger-1500 (blue dashed line) and TA-1200 (red dashed line).}
\label{fig:biasAndResolution}
\end{figure}

To illustrate this impact of this on the spectrum, we have modeled the biases and resolutions presented here to quantify the resulting expected differences in flux.
This has been applied to the spectrum reconstructed by the Pierre Auger Observatory~\cite{AugerSpectrum2020} before performing a comparison with the spectrum of the Telescope Array experiment~\cite{TASpectrum2019}.
In the case of Auger-1500, the parametrisation of the bias and resolution are extracted from \cite{SDRec2020} and shown (in blue) in Fig.~\ref{fig:biasAndResolution}. As adopted by the Pierre Auger Collaboration, the bias and resolution are defined for showers at the median value of the distribution of the zenith, 38$^\circ$.
For the simulated set of TA-1200, the bias and resolution are extracted from the studies performed in the previous sections.
To reproduce the procedure applied to the events recorded by TA, in the TA-1200 simulated events, the closest detector is assumed to saturate~\cite{IvanovThesis2012} and not used in the reconstruction of the shower size.
In addition to the statistical uncertainties that are decreasing when the energy of the primary particle increases, and with it the multiplicity of detectors used in the reconstruction of the LDF, the uncertainties coming from the choice of a non-optimal distance at 800$\,$m are reported and explain the degradation of the resolution with energy.
The results are shown in Fig.~\ref{fig:biasAndResolution}-left (in red). 

The bias in the estimator of the shower size, then propagated to the energy, is roughly estimated from Fig.~\ref{fig:biasAndResolution}-right in the worst case presented and modeled by a cubic function yielding a logarithmic increase of 20\% per decade in energy.
This result is in accordance with the comparison of Auger and TA spectra performed by the two collaborations~\cite{SpectrumWG2021}.

\begin{figure}[t]
    \centering
    \includegraphics[width=0.75\textwidth]{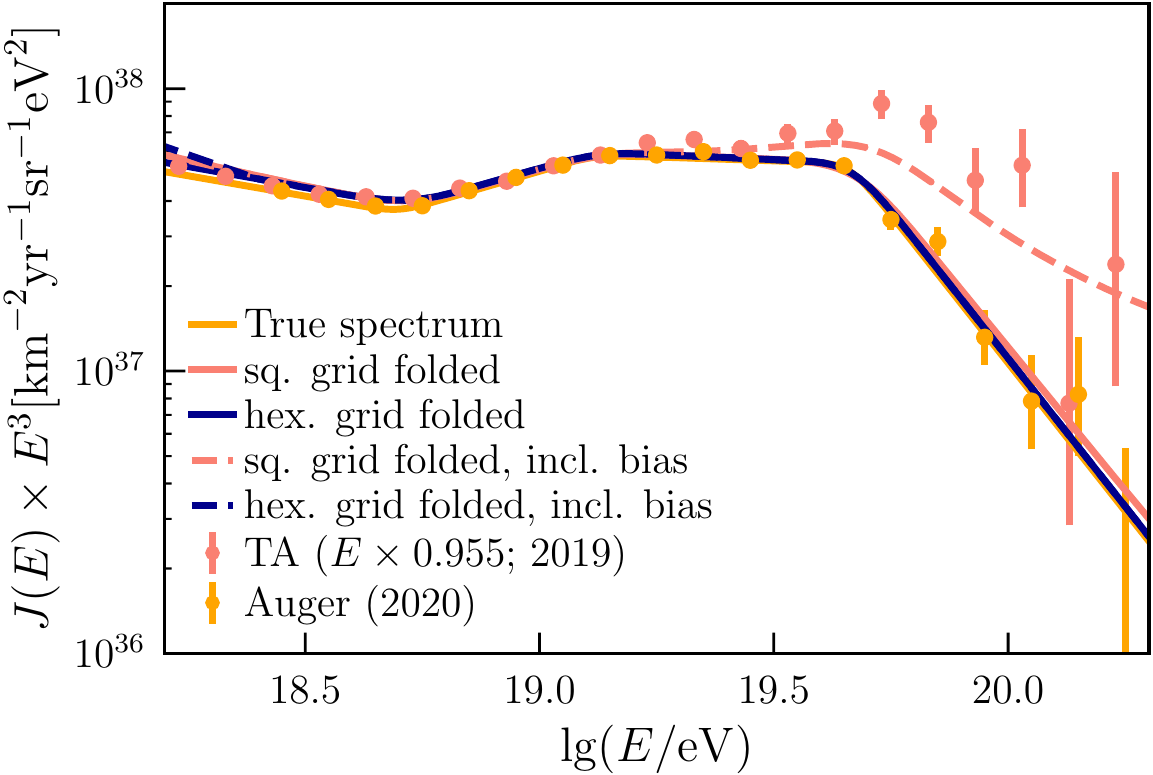}
    \caption{Energy spectrum of the underlying spectrum (true spectrum) and the folded spectra of the TA-1200 and Auger-1500 grid for two cases: with and without accounting for the additional bias from Fig.~\ref{fig:biasAndResolution}. For reference, Auger~\cite{AugerSpectrum2020} and TA~\cite{TASpectrum2019} spectra are shown.}
    \label{fig:spectra}
\end{figure}

The deconvolved spectrum reported by the Pierre Auger Collaboration is taken as the reference to illustrate the impact of the bias and resolution extracted in our study. 
The bias and resolution from Auger-1500 and TA-1200 are applied to this spectrum and shown in Fig.~\ref{fig:spectra}. 
Applying the resolution only, the resulting folded spectra do not deviate more than 8-10\% from the reference spectrum.
In the case of Auger-1500, the bias reported is null except at the lowest energies and the deviation is below 10\%. 
Now, considering the bias from TA-1200, the flux at the highest energies is more than 3 times larger than the original flux.
Whether we are looking at the Auger-1500 or TA-1200 modified spectra, the bias is the crucial point in this study while the resolution has a relatively minor impact. 
Thus, any bias emerging from non-linearities in the energy estimator, introduced because of a lack of knowledge of the true LDF, will directly impact the spectrum.

On top of the folded spectra in Fig.\ref{fig:spectra}, the measured spectra observed by Auger and TA are represented by the blue and red points respectively. 
The spectrum of TA has been rescaled to give an agreement between the two spectra in the ankle region.
Part of the remaining differences observed between the two spectra, i.e. the 10\% systematic shift per decade, could be explained by the bias observed in the study performed with the TA-1200 simulated set.

\section{Conclusion}
Studying the cosmic rays at the highest energies requires larger and larger array of detectors but at the cost of an increase of the distance between two adjacent detectors. 
Thus, it requires the choice of a distance at which fluctuations of the density of particles are minimal, to compute the shower size and get an accurate estimator of the energy.
The results reported above show that the choice of array geometry is important when selecting the optimal distance, $r_\text{opt}$, at which the signal size to be related to shower energy is measured. 
In the case of the layout of the Auger Observatory, the choice made is independent of zenith angle and primary energy. 
However, this is not the case for the Telescope Array where dependencies of $r_\text{opt}$ on energy and zenith angle have been found. 
The choice of a non-optimal distance to estimate the shower size leads to non-linearities in the determination of the energy causing the introduction of a bias affecting the spectrum. 
In the \emph{worst} case explored here, a change of the slope of the LDF by $-1\sigma$, this distortion of the spectrum could contribute to the differences observed between the spectra of Auger and TA. 

The cause of the energy and zenith dependencies are still under investigation, whether it comes from the layout of the array, the choice of the LDF to reconstruct the lateral profile, etc. and are the subject of discussions between the Pierre Auger and Telescope Array Collaborations.
With the current generations of surface detectors dedicated to the detection of UHECRs entering their second phase, the next detector has to be planned. 
It is thus desirable to get an estimator of the energy as accurate as possible. 
Therefore a careful study of the optimal distance and the LDF will have to be taken into account when defining the detectors to be built.


\end{document}